\documentclass[a4paper,11pt]{article}

  \title{Jeffreys Priors versus Experienced Physicist Priors\\
Arguments against Objective Bayesian Theory}

\author{Giulio D'Agostini \\
Universit\`a di Roma ``La Sapienza'' and INFN, 
Rome (Italy)}

\date{}

\begin{document}
\maketitle
\begin{abstract}

\footnotetext{Email: dagostini@roma1.infn.it.\
URL: http://www-zeus.roma1.infn.it/$^\sim$agostini/
\\Contributed paper to the 6th Valencia International 
Meeting on Bayesian Statistics, Alcossebre (Spain), 
May 30th - June 4th, 1998.}
I review the problem of the choice of the priors from 
the point of view of a physicist interested in 
measuring a physical quantity, and I try to show that 
the reference priors often recommended for the purpose 
(Jeffreys priors) do not fit to the problem. Although 
it may seem surprising, it is easier for an ``experienced
physicist'' to accept subjective priors, or even 
purely subjective elicitation of probabilities, without 
explicit use of the Bayes' theorem. The problem 
of the use of reference priors is set in the more general 
context of  ``Bayesian dogmatism'', which could really 
harm  Bayesianism. 
\end{abstract}

\section{Introduction}
The choice of the prior is usually felt to be a vital
problem by all those who approach the Bayesian methods 
with a purely utilitarian spirit, without having assimilated 
the philosophy of subjective probability. Some 
use ``Bayesian formulae'' simply because they 
``have proved'', by Monte Carlo simulation, that they 
work in a particular problem. Others
like the principles of Bayesian reasoning, but are 
embarrassed by the apparent ``arbitrariness'' 
of the priors. Just to mention an example of this second attitude, 
I have been told of a conference\cite{Pia}
 in which 
astrophysicists discussing which statistical methods to use
concluded, more or less:
{\it ``yes, Bayesian statistics looks nice, but now
we should make an effort to define our priors in an objective way''}. 
The use of  the reference 
priors (hereafter I will refer only to  
Jeffreys' priors\cite{Jeffreys}, the most common in Physics 
applications) gives 
a chance to avoid taking responsibility when assessing which 
priors are suitable for a specific problem, and it gives 
the {\it illusion of objectivity} (the dream of the
simple minded practitioner). Although I agree 
on the validity of  a
{\it  ``concept of a `minimal informative' prior specification
- appropriately defined!''}\cite{BS}, 
to those who are not fully aware 
of the intentions and limits of  reference analysis, 
the Bayesian approach can be perceived as dogmatic.
In this paper I would like to comment on Jeffreys' priors
from the side of the ``experienced physicist''\footnote{With 
this generic name I mean whoever is used to an everyday 
confrontation with real data.}, a point of view often 
neglected, since this matter is more often debated 
among mathematicians, statisticians or philosophers.  
So, instead of focusing on  the original intentions of 
Jeffreys' priors, I will  criticize their uncritical use, 
the shadow of dogmatism they diffuse on the theory, and 
the unstated psychological motivations of some 
of their supporters.   
In contrast, I will stress 
the guiding 
role of the {\it guess} of the ``expert'', which allows the 
subjective 
assessment of uncertainty even in the case of a single observation. 
To make this point understandable to those who 
are not familiar with experimentation, I will give a brief 
reminder, later, 
of how measurements are actually performed 
and I will comment on 
the ISO (International 
Organization for Standardization)
recommendations concerning  measurement uncertainty. 

\section{Bayesian dogmatism and its dangers}
In principle there is little to comment on 
the indiscriminate use and uncritical recommendation of 
reference priors. It is enough to glance at 
many books, lecture notes, articles and conference proceedings 
on Bayesian theory and applications. I would just like to 
give  an example which concerns me very much, because it may 
influence the 
High Energy Physics
community 
to which I belong. In a paper which recently appeared in   
{\it Physical Review}\cite{CF} 
 it is stated that 
\begin{quote}
{\small 
{\it ``For a parameter  $\mu$ which is restricted to $[0,\infty]$, 
a common non-informative prior in the statistical literature is
$P(\mu_t)=1/\mu_t$\ldots In contrast the 
PDG}\footnote{PDG stands for ``Particle Data Group'', 
a committee that publishes every second year the {\it Review 
of Particle Properties}\cite{PDG}, a very influential collection
of data, formulae and methods, including sections on 
Probability and Statistics.}{\it 
description is equivalent to using a prior which is uniform 
in $\mu_t$. This prior has no basis that we know of in Bayesian
theory''} 
}
\end{quote}
This example should be taken really very seriously. The authors
in fact use the pulpit of a prestigious 
journal\footnote{This is also an example of  bad 
style, publishing a paper in a Physics journal, 
pretending that it is a contribution to statistical theory,
but avoiding undergoing the scrutiny of
a more appropriate referee
({\it ``In this paper, we use the freedom inherent in Neyman's
construction in a novel way to obtain a unified set of classical
confidence intervals for setting limits and quoting two-sided confidence
intervals. The new element is a particular choice of ordering, 
based on likelihood ratios, which we substitute for more common 
choices in Neyman' construction''}\cite{CF}.) 
.}
 to appear 
as if they understand 
 well both the Bayesian and the classical 
approach and, on this basis, they discourage the use of Bayesian methods 
({\it ``We then obtain confidence intervals which are never 
unphysical or empty. Thus they remove an original intention
for the description of Bayesian intervals 
by the PDG''}).

So, while someone can be in favour of default use of reference priors, 
which may have some advantage in 
attracting practitioners reluctant to subjectivism, it seems 
to me that in the long term it can play against the Bayesian 
theory, in a similar way to that which happened at the end of 
last century, because of the abuse of  uniform distribution. 
This worry is well expressed in John Earman's  conclusions 
to his 
``critical examination
of Bayesian confirmation theory''\cite{Earman}:
\begin{quote}
{\small\it ``We than seem to be faced with a dilemma. 
On the one hand, Bayesian considerations seem 
indispensable in formulating and evaluating scientific inference. 
But on the other hand, the use of the full
Bayesian apparatus seems to commit the user to a form of 
dogmatism''.}
\end{quote}
\section{Unstated psychological motivations behind Jeffreys' priors?}
From the most general (and abstract) point of view, it 
is not difficult 
to agree that ``in one-dimensional continuous regular problems,
Jeffreys' prior is appropriate''\cite{BS}. Unfortunately, it is
rarely the case that in physical situations
the status of prior knowledge 
is equivalent to that expressed by the Jeffreys' priors, as I will 
discuss later. Reading ``between the lines'', 
it seems to me that the reason for choosing them is essentially 
psychological. For instance, when used to infer $\mu$ (typically 
associated with the ``true value'') from
``Gaussian small samples'', the use of a prior of the kind 
$f_\circ(\mu,\sigma)\propto 1/\sigma$ has two {\it formal benefits}:
\begin{itemize}
\item
first, the mathematical solution is simple (this reminds 
me of the story of the drunk under the streetlamp, 
looking for the key lost in the dark alley);
\item 
second, one recovers the Student distribution, and for some 
it seems to be reassuring that a Bayesian result gets blessed by 
{\it ``well established''} frequentistic methods. 
(``We know that this is the right solution'', 
a convinced Bayesian once told me\ldots)
\end{itemize}
But these 
arguments, never explicitly stated, cannot be accepted, 
for obvious reasons. I would like only to comment on the 
Student distribution, 
the ``standard way'' for 
handling small samples, although  
 there is in fact no deep reason 
for aiming to get such a distribution for the posterior. 
This becomes clear to 
anyone who, having measured  the size of 
this page twice and having found a difference 
of 0.3 mm between the measurements, then has to base  
his conclusion on that distribution. 
Any
rational person  
will refuse to state that, in order 
to be 99.9\,\% confident in the result,
the uncertainty interval should be 9.5 cm wide
(any carpenter would laugh\ldots). This may be the
reason why, as far as I know, physicists don't
use the Student distribution.  

Another typical application of the Jeffrey' prior is in the 
case of inference on the $\lambda$ parameter 
of a Poisson distribution, having observed a 
certain number of events.
Many have, in fact, a reluctance
to accept as an estimate of $\lambda$ a value which differs from 
the observed number of counts 
(for example, $\mbox{E}(\lambda)=x+1$ starting from a uniform
prior) and which is deemed to be distorted 
by the ``distorted'' frequentistic criteria to analyse the problem. 
In my opinion, in this case one should simply
 educate the practitioners about the difference
between the concept of maximum belief and that of prevision
(or expected value).  An example in which the difference becomes
 crucial is the case where no counts are observed, 
a typical situation for frontier physics, 
where new phenomena are constantly looked for. Any reasonable prior
consistent with an investigated rare process, close to the  
limit of experimental 
sensitivity, provides reasonable results (even a uniform prior
is good for the purpose) and allows the calculation of 
 ``upper limits''. Instead, a prior of the kind 
$f_\circ(\lambda)\propto1/\lambda$
prevents the use of any quantitative probabilistic statement to
summarize the achievement of the measurement
and the same result ($0\pm0$) 
will come out independently of the size, sensitivity and running
time of the experiment.

In the following I will only consider the case of normally 
distributed observations. 

\section{Unavoidable prior knowledge behind any measurement}
To understand why an ``experienced physicist'' has difficulty in
 accepting 
a prior of the kind $f_\circ(\sigma) \propto 1/\sigma$
(or $f_\circ(\ln(\sigma)) = k$), one has to remember
 that the process 
of measurement is very complex (even in everyday situations, 
like measuring the size of the page 
\underline{You} are reading 
\underline{now}, 
just to avoid abstract problems): 
\begin{itemize}
\item
 first You have to {\it define the measurand} (the quantity 
 we are interested in);
\item
 then You have to {\it choose the appropriate instrument}, 
having known properties,  well suited range and resolution, 
and in which You have some confidence, achieved on the basis of previous
measurements;
\item 
 the {\it measurement} is performed and, if possible, 
repeated several times;
\item 
then, if needed, You apply  {\it corrections}, 
also based on previous experience with that kind of
measurement, in order to take into account
known (within uncertainty) systematic effects; 
\item 
finally\footnote{This is not really the end of the story 
if You wish 
Your result to have some impact on the scientific community
(or simply on commerce).
Only if other people trust You, will they
use the result in further scientific (or business)
reasoning, as if it were their own result.} 
You get a credibility interval for the quantity
(usually a {\it best estimate} with a related {\it uncertainty});
\end{itemize}
Each step involves some prior knowledge and, typically,
each person who performs the measurement 
(either a physicist, a biologist, a carpenter or a bricklayer)
operates in his field of expertise. This means that he
is   well aware of the 
error he might make, and then of the uncertainty
associated with the result. This is also true if only 
a single observation has been 
performed\footnote{This defence of the possibility 
of quoting an uncertainty from a single measurement 
has nothing to do with 
the mathematical games like those of \cite{Rodriguez}.}:
try to ask 
a carpenter how much he believes in his result, 
possibly helping him to quantify the uncertainty 
using the concept of the coherent bet. 

There is also another important aspect of the 
``single measurement''. One should  
note that many measurements, which seem to be due to a single observation, 
consist in fact of several observations 
made within a short time: for example, 
measuring a length with a design ruler, one checks several times 
the alignment of the zero mark with the beginning 
of the segment to be measured;
or, measuring a voltage with a voltmeter or 
a mass with a balance, one waits until the reading 
is well stabilized. Experts use unconsciously 
information of this kind when they have to state an uncertainty.

The fact that the evaluation of uncertainty does not 
come necessarily from repeated measurements has also been 
recognized by the International Organization for Standardization
(ISO)
in its {\it ``Guide to the expression of uncertainty in 
measurement''}\cite{ISO}. 
There the uncertainty
is classified {\it ``into two categories according to the way their 
numerical value is estimated:
\begin{enumerate}
\item[A.] those which are evaluated by 
          statistical methods\footnote{Here ``statistical'' 
stands for
``repeated observations on the same measurand.''};
\item[B.] those which are evaluated by other means;''\cite{ISO}  
\end{enumerate}
}
Then, illustrating the ways to evaluate the  
``type B standard uncertainty'', the {\it Guide} states that
\begin{quote}
{\small \it ``the associated estimated variance $u^2(x_i)$ or the standard
uncertainty $u(x_i)$ is evaluated by scientific judgement based on all
of the available information on the possible variability of $X_i$. 
The pool of information may include
\begin{itemize}
\item[-] previous measurement data; 
\item[-] experience with or general knowledge of the behaviour
and properties of relevant materials and instruments;
\item[-] manufacturer's specifications;
\item[-] data provided in calibration and other certificates;
\item[-] uncertainties assigned to reference data taken from handbooks.''
\end{itemize}
}
\end{quote}
It is easy to see that the above statements have sense
only if the probability is interpreted as degree of belief,
as explicitly recognized by the {\it Guide}:
\begin{quote}
{\small \it
``\ldots Type B standard uncertainty is obtained from an 
assumed probability density function based on the degree of belief that 
an event will occur [often called subjective probability\ldots].'' 
}
\end{quote}
It is also interesting to read the
worries of the {\it Guide} concerning the uncritical use of 
statistical methods and of abstract formulae:
\begin{quote}
{\small\it ``the evaluation 
of uncertainty is neither a routine task nor a 
purely mathematical one; it depends on detailed knowledge
of the nature of the measurand and of the measurement.
The quality and utility of the uncertainty quoted for
the result of a measurement therefore ultimately 
depend on the understanding, critical analysis, 
and integrity of those who contribute to the assignment 
of its value''\cite{ISO}.
}
\end{quote}
This appears to me perfectly in line with the lesson
of  genuine subjectivism, accompanied by the 
normative rule of  coherence\cite{definetti}. 
It is instead surprising to see how many Bayesians
seek refuge in stereotyped formulae or to see how many  
still stick to the frequentistic idea that repeated observations
are needed in order to evaluate the uncertainty of a measurement. 

\section{Rough modelling of realistic priors}
After these comments on measurement, it becomes clearer why 
a prior of the kind $f_\circ(\mu,\sigma)\propto 1/\sigma$
does not look natural.  
As far as $\sigma$ is concerned, this prior would imply that 
standard deviations ranging over  several orders of magnitude 
would be equally possible. This is unreasonable in most cases. 
For  example, measuring the size of this
page, no one would expect  $\sigma\approx {\cal O}(1\,\mbox{cm})$ or
$\approx {\cal O}(1\,\mu\mbox{m})$. Coming to $\mu$, the choice
$f_\circ(\mu)=k$ is acceptable until $\sigma\ll \mu$ (the 
so called Savage {\it principle of precise 
measurement}\cite{Savage}). But when the order of magnitude 
of $\sigma$ is uncertain, the prior on $\mu$  
should also be revised (for example, most of the directly measured
quantities are positively defined).  

Some priors which, in my experience, are closer 
to the typical prior knowledge of the person 
who makes {\it routine measurements} 
are those concerning the order of magnitude of $\sigma$, or the order
of magnitude on the precision (quantified by the 
variation coefficient $v=\sigma/|\mu|$).  
For example\footnote{For sake of simplicity, let us stick to the case 
in which the fluctuations are larger that the intrinsic 
instrumental resolution. Otherwise one needs to model 
the prior (and the likelihood) with a discrete distribution.},
one may expect a r.m.s. error of 1 mm, but 
values of  0.5 or 2.0 mm would not look surprising. Even 
0.1 or 2 mm would look possible, but certainly not
10\,$\mu$m or 2 cm. Alternatively, for other measurements,
what matters could be the order of magnitude of the class of 
precision. In both cases 
a distribution which seems suitable for a rough modelling 
of this kind of priors is a {\it lognormal} in either $\sigma$ 
or $v$. For instance, the above example could be modeled 
with $\ln{\sigma}$ normally distributed with average 0
 ($=\ln 1$) and standard deviation 0.4. The 1, 2 and 3 standard 
deviation interval on $\sigma$/mm would be 
$[0.7, 1.5]$, $[0.5, 2.2]$ and $[0.3, 3.3]$, respectively, 
in qualitative agreement with the prior knowledge. 

In the case of more sophisticated measurements in which the 
measurand is a positive defined quantity 
 of unknown order of magnitude a suitable prior of $\mu$ 
is flat in $\ln \mu$ (before the first measurement you 
don't know the order of magnitude you will get), while 
of $\sigma$ is somehow correlated to $\mu$ 
($v$ is expected, reasonably, to lie in a range, the extremes of which 
do not differ by too many orders of magnitudes).

One may think of other possible measurements which give rise 
to other priors, but I find it very difficult to imagine 
a real situation for which the Jeffrey's priors are 
appropriate.

\section{Purely subjective assessments}
In the previous section I have given some suggestions for
solving the problem within the framework of the Bayes' theorem
paradigm. But I don't want to give the impression
that this is the only way to proceed.

The most important teaching of subjective probability 
is that probability is always conditioned by a given status 
of information. The probability is updated in the light 
of  any new information. But it is not always possible to 
describe the updating mechanism using the neat scheme of 
the Bayes' theorem. This is well known in many fields, 
and, in principle, there is no reason  for considering the 
use of the Bayes theorem to be indispensable to 
assessing uncertainty in scientific measurements. The idea is to force the 
expert to declare (using the coherent bet) some quantiles 
in which he believes is contained the true value, on the basis
of a few observations. It may be easier for
him to estimate the uncertainty in this way, drawing on his 
past experience, rather than trying to model some priors and to play with 
the Bayes' theorem. The message is what experimentalists
intuitively do: {\it when you have just a few observations,
what you already know is more important than what 
the standard deviation of the data teaches you}.

Some will probably be worried by the arbitrariness of this conclusion,
but it has to be remembered that: an expert 
can make very good guesses in his field;  20, 30, or even 50 \%
uncertainty in the uncertainty is not considered to 
significantly spoil the quality of a measurement; there 
are usually many other sources of uncertainty, due to 
possible systematic effects on unknown size, which 
can easily be more critical. 
I am much more worried by the attitude of giving up
prior knowledge to a mathematical convenience, since this 
can sometimes lead to paradoxical results. 

\section{Conclusions}
The default use of Jeffreys priors is clearly unjustified, 
especially in inferring the parameters of the 
normal distribution, the model mainly used in physics 
measurements. A more realistic choice of the priors 
would be lognormal in $\sigma$ or in the variation coefficient, 
but the posteriors do not have a closed form and nobody 
wants to make complicated calculations in routine measurements. 
When the number of measurements is of the order of the 
unit it can be more reasonable to use just subjective
estimates in the light of the observed data and of  
past experience. This corresponds to the practice
of ``experienced physicists'', who tend to trust more in 
prior experience when they are not able to perform many 
measurements. In particular, it is absolutely legitimate 
to state the uncertainty, even if only a single measurement has
been made, when one has the appropriate prior knowledge. 
This has also been recognized by the metrological authorities. 

As a more general remark, 
I find all attempts to put the Bayesian theory on dogmatic grounds very
dangerous.  Not only because this can sometimes lead to absurd results in 
critical situations, but also because such results can seriously damage
 the credibility of the Bayesian theory itself.


\begin{thebibliography}{ref99}

\bibitem{Pia}
Conference on {\it ``Statistical Challenges in Modern Astronomy II''}, 
The Pennsylvania State University, University Park, Pennsylvania, U.S.A.
June 2 - 5, 1996 (private communication by P. Astone). 

\bibitem{Jeffreys}
H. Jeffreys, {\it ``Theory of probability''}, Oxford
University Press, 1961.

\bibitem{BS}
J.M. Bernardo, A.F.M. Smith,
``{\it Bayesian theory}'', John Wiley \& Sons Ltd, Chichester, 1994.

\bibitem{CF}
G.J. Feldman and R.D. Cousins, 
{\it ``Unified approach to the classical statistical 
analysis of small signals''}, {\it Phys. Rev. D} {\bf 57} (1998) 3873.

\bibitem{PDG}
Particle Data Group, R.M. Barnet et al., 
{\it ``Review of particle properties''},
{\it Phys. Rev. D} {\bf 54} (1996) 1.

\bibitem{Earman}
J. Earman, {\it ``Bayes or bust? A critical examination of
Bayesian confirmation theory''}, The MIT Press, 1992. 

\bibitem{Rodriguez}
C.C. Rodriguez, {\it ``Confidence intervals from one observation''},
unpublished 
(paper available in http://omega.albany.edu:8008/)

\bibitem{ISO}
International Organization for Standardization (ISO),
{\it ``Guide to the expression of uncertainty in measurement''},
Geneva, Switzerland, 1993.

\bibitem{definetti}
B. de Finetti, {\it ``Theory of probability''},
J. Wiley \& Sons, 1974.

\bibitem{Savage}
L.J. Savage et al., 
{\it ``The foundations of statistical inference: a discussion''},
Methuen, London, 1962.

\end{thebibliography}
\end{document}